# Generation, Implementation and Appraisal of an N-gram based Stemming Algorithm

**B. P. Pande, Pawan Tamta, H. S. Dhami**

## Abstract

A language independent stemmer has always been looked for. Single N-gram tokenization technique works well, however, it often generates stems that start with intermediate characters, rather than initial ones. We present a novel technique that takes the concept of N-gram stemming one step ahead and compare our method with an established algorithm in the field, Porter's Stemmer. Results indicate that our N-gram stemmer is not inferior to Porter's linguistic stemmer.

## 1. Introduction:

Stemming, a very useful tool under the domain of Information Retrieval (IR), supported by almost each and every modern indexing and search systems. Sometimes put under a single roof with its sibling process, lemmatization, stemming is nothing but to reduce the morphological variants of a word into a single form- the stem. Before indexing, truncating attached affixes (suffixes or prefixes) from terms, makes thus generated stem to represent a broader concept than the original term, the stemming process eventually increases the number of retrieved documents in an IR system. Although, the effectiveness of stemmers has always been argued [7, 8], still stemming is kept as an active method in the IR toolbox.

We have a rich literature of stemming process- there are linguistic and statistical approaches. The former are *rule based* techniques [4, 8, 11, 15, 17], in which a priori knowledge of the target language morphology is required. The latter are subject to exhaustive statistical analysis of the corpus, but work in a language-neutral way [5, 6, 12-14, 19]. There are also some lexicon based approaches [9]. Linguistic stemmers have one major disadvantage that they are not suitable at multilingual environment. Statistical algorithms, such as one which is based on character frequencies can cope up with this limitation.

In the urn of the statistical techniques, the most interesting and robust technique that caught our attention is *Single N-gram Stemming* by James Mayfield and Paul McNamee [13]. The authors hypothesized that exploiting overlapping sequences of *N* characters, the benefits of stemming can be achieved in a language independent way. They argued that some of the N-grams (actually, low frequent) derived from a word will be common to only those portions of the word that do not exhibit morphological variation, i.e. they can serve as an adequate stem substitute. They suggested selecting the word internal N-gram with the highest inverse document frequency (IDF) as a word's *pseudo stem.* They gave an excellent reasoning that a particular morphological variation being exhibited by affixes tends to be repeated across many different words, and therefore exhibit low IDF. As an instance, they tabled document frequencies of different N-grams of the word 'juggling' (from the CLEF 2002 collection) and the suggested pseudo-4 and pseudo-5 stems, according to their concept.

The key behind our present research resides in the tables being talked above. The pseudo-4 stem for the words currency and warrens were *rren* and *rens* respectively. The pseudo-5 stem for juggler, currency and warrens were *ggler, rency and rrens* respectively. A general insight says that a stem has to start from the beginning character of the word, not from anywhere else. So, is there any scope of improvement over this already elegant method?

Endeavouring to answer the above question is the essence of our present paper. We work for the better utilization of N-grams, keeping the idea of counting their frequencies intact. Rather than depending on the simple concept of IDF, we develop a novel algorithm that gives better approximations of stems.

## 2. Proposed Method:

The authors [13] reported in their experimental results that out of the raw words, Snowball stems [18], pseudo stems and 4-grams, the trend was for 4-grams to perform best. They concluded the fact that 4-grams, the initial 4 characters of any word are the best representation of a word stem. This strengthens our insight that for a valid (not from a strict linguistic point of view) and nicely approximated stem, it should start from the beginning. So what if instead of finding frequencies of overlapping N characters (i.e. words internal N-grams), we approximate stem from the frequencies of 4-gram, 5-gram, 6-gram and so on? Seems a promising direction.

Our idea is to take 4-gram as our initial guess for finding the stem. Obviously, this rules out words with length 3 or less. This would be a stunning advantage as many of the stop words are of length 3 or less. Now, for a given word, what we are left with finding frequencies of 4-gram, 5-gram and so on up to the length of the word and selecting the best from them. Before we embark upon our approach, let us look at some of the different N-gram frequencies from COCA [3]. We start with the same word *juggling*[1].

Table 2.1: Different N-gram frequencies from COCA for the word 'juggling'

| ***N-gram*** | ***Frequency*** | ***N-gram*** | ***Frequency*** |
|---|---|---|---|
| jugg* | 915 | jugglin* | 328 |
| juggl* | 729 | juggling* | 328 |
| juggli* | 328 | | |

We infer that an N-gram of a word that serves as an adequate approximation of the stem must has a relative good frequency across the corpus. All the N-grams less than this must be highly frequent and above this tend to exhibit lower frequencies gradually, ending in the total frequency of the given word (there may be the case where two or more consecutive entries are identical.)Well, above table successfully meets our assumption.

Thus, we have an integer variable in hand whose values tend to decrease (or remain constant) with the increase in the length of a partial word (N-gram) by one unit. It is very likely that this fall is very steep at the transition from invariant part of a word (stem) into the derivational or suffix part. But how can we catch a relative maximum fall? One way is to take deviation of each two consecutive entries. We define a metric as under

$$\lambda_i = \begin{cases} abs(F_i - F_{i-1}) & if\ i > 4 \\ 0 & else \end{cases} \qquad (2.1)$$

[1] The NEWSPAPER search option was selected and the minimum frequency was set to 10.

Where $F_i$ is the frequency of the i[th] N-gram. Next we calculate the second order deviation

$$\Delta_i = \lambda_i - \lambda_{i-1} \tag{2.2}$$

Let $\psi_N$ be the variable corresponding to the N-gram of length *N*. We define our procedure in two phases:

Phase 1:

1) Initialization:

    $\lambda_4 = M$ where *M* is a very high integer value

    $\psi_N = 4$

2) Recursion:

    Calculate $\lambda_i$, 4< i ≤ |word|

    If $\lambda_i > \Gamma$

    $$\psi_N = \underset{4<i\le|word|}{argmax}[F_i, F_{i-1}]$$

    Else

    $$\psi_N = i$$

3) Termination:

    a.) If N=|word|, stop. Else, calculate $\Delta_i$.

    b.) If $\Delta_i$>0, stop.

Where $\Gamma$ is a threshold value, which defines how much frequency deviation of two N-grams is acceptable. It is kept either zero or to a minimum value. Recursion step does the looping through the increasing N-grams of the input word. If at the end of Phase 1, $\psi_N = |W|$ , move to Phase2.

Phase 2:

If the frequencies of last three N-grams ($F_i$ s) are identical, delete the last three characters of the input word only if deleting them makes the resulting stem greater than three, i.e.

$$\psi_N = \psi_N - 3 \qquad \text{If } \psi_N - 3 > 3$$

At the end of the above procedure, $\psi_N$ contains the value of the N-gram that approximates the required stem. Phase 2 controls over that words that exhibit common suffixes (like *–ing*), if they aren’t stripped in Phase 1.

## 3. Illustrations and Experiments:

Let us apply our method to the cluster of English, Spanish and Portuguese words. What we need is sequential frequencies of N-grams from a rich corpus. For English language, we relied on COCA [3]. This corpus has two benefits, first it has a rich collection of English words and second, it is easy to calculate the N-gram frequencies using wild card [*]. Sequential frequencies for Spanish and Portuguese words are taken from Corpus Del Español [1] and Corpus Do Português [2] respectively. The results over some testing candidates are being tabled under:

Table 3.1: Word clusters and stems

| Cluster no. | Word | Stem | Cluster no. | Word | Stem |
|---|---|---|---|---|---|
| 1 (English) | create | creat | 2 (Spanish) | trabajan | traba |
| | creates | creat | | trabajar | traba |
| | creating | creat | | trabajado | traba |
| | created | creat | | trabajador | traba |
| | creation | creat | 3 (Portuguese) | dificil | dific |
| | creative | creat | | dificilmente | dific |

Now, let us compare our stemmer empirically with an established linguistic stemmer, say Porter's stemmer [17]. We select English as our testing language as outputs of Porter's algorithm can be easily extracted at Snowball project [18]. We take sample of 100 random English words and apply Porter's Snowball framework and our procedure over them. The results are reported in Table 3.2. Identical stems generated by either algorithm are shown in bold. Four POS, Noun, Adjective, Verb and Adverb were taken into account. We select *Newspaper* search area in COCA.

To compare the stemming results of Snowball (Porter) stemmer and our N-gram stemmer, we employed direct evaluation method. Like the comparison made by Chris D. Paise [16], using length truncation as a baseline, we employed *Levenshtein distance* [10] as a base measure. The distance is the number of deletions, insertions, or substitutions required to transform the source string into the target string. The Levenshtein distance therefore, represents by how many units a word has been stripped by a stemmer. For our sample of the 100 words, these distances are shown in Table 3.2, $LD_{Snow}$ and $LD_{N\text{-}gram}$ for Snowball stems and our N-gram stems respectively.

Table 3.2: Random Words and their stems

| S. No. | Word | Snowball stem | N-gram stem | $LD_{Snow}$ | $LD_{N\text{-}gram}$ |
|---|---|---|---|---|---|
| 1 | Parsons | **Parson** | **Parson** | 1 | 1 |
| 2 | Dilution | **Dilut** | **Dilut** | 3 | 3 |
| 3 | Agreement | Agreement | Agree | 0 | 4 |
| 4 | Passion | **Passion** | **Passion** | 0 | 0 |
| 5 | Cutter | **Cutter** | **Cutter** | 0 | 0 |
| 6 | Refill | Refill | Refil | 0 | 1 |
| 7 | Museums | **Museum** | **Museum** | 1 | 1 |
| 8 | Stallion | **Stallion** | **Stallion** | 0 | 0 |
| 9 | Braid | Braid | Brai | 0 | 1 |
| 10 | Fleet | Fleet | Flee | 0 | 1 |
| 11 | Midwife | Midwif | Midwife | 1 | 0 |
| 12 | Haze | **Haze** | **Haze** | 0 | 0 |
| 13 | Prophet | Prophet | Prophe | 0 | 1 |
| 14 | Vacancy | **Vacanc** | **Vacanc** | 1 | 1 |
| 15 | Ninety | Nineti | Ninet | 1 | 1 |
| 16 | Menace | **Menac** | **Menac** | 1 | 1 |
| 17 | Laceration | Lacer | Lacerat | 5 | 3 |
| 18 | Training | **Train** | **Train** | 3 | 3 |
| 19 | Librarian | Librarian | Librar | 0 | 3 |
| 20 | Ordaining | **Ordain** | **Ordain** | 3 | 3 |
| 21 | Mainstream | Mainstream | Mainst | 0 | 4 |
| 22 | Bloodier | Bloodier | Blood | 0 | 3 |
| 23 | Abject | **Abject** | **Abject** | 0 | 0 |
| 24 | Substitute | **Substitut** | **Substitut** | 1 | 1 |
| 25 | Overseas | Oversea | Overseas | 1 | 0 |
| 26 | Freehand | Freehand | Freehan | 0 | 1 |
| 27 | Despotic | Despot | Despoti | 2 | 1 |
| 28 | Predicted | Predict | Predic | 2 | 3 |
| 29 | Lyric | **Lyric** | **Lyric** | 0 | 0 |
| 30 | Eleventh | Eleventh | Eleven | 0 | 2 |
| 31 | Admonishing | Admonish | Admoni | 3 | 5 |
| 32 | Quiet | **Quiet** | **Quiet** | 0 | 0 |
| 33 | Macabre | Macabr | Maca | 1 | 3 |
| 34 | Unloaded | **Unload** | **Unload** | 2 | 2 |
| 35 | Quizzical | Quizzic | Quizzi | 2 | 3 |
| 36 | Alto | **Alto** | **Alto** | 0 | 0 |
| 37 | Undeveloped | Undevelop | Undevelo | 2 | 3 |
| 38 | Casual | **Casual** | **Casual** | 0 | 0 |
| 39 | Recognized | Recogn | Recogni | 4 | 3 |
| 40 | Devastating | Devast | Devastat | 5 | 3 |
| 41 | Abused | Abus | Abuse | 2 | 1 |
| 42 | Abusing | Abus | Abusing | 3 | 0 |
| 43 | Admired | Admir | Admire | 2 | 1 |
| 44 | Admiring | **Admir** | **Admir** | 3 | 3 |
| 45 | Believed | Believ | Belie | 2 | 3 |
| 46 | Believing | Believ | Belie | 3 | 4 |
| 47 | Borrowed | **Borrow** | **Borrow** | 2 | 2 |
| 48 | Borrowing | **Borrow** | **Borrow** | 3 | 3 |
| 49 | Carried | Carri | Carrie | 2 | 1 |
| 50 | Carrying | Carri | Carry | 3 | 3 |
| 51 | Consulted | **Consult** | **Consult** | 2 | 2 |
| 52 | Consulting | **Consult** | **Consult** | 3 | 3 |
| 53 | Deceived | Deceiv | Deceive | 2 | 1 |
| 54 | Deceiving | **Deceiv** | **Deceiv** | 3 | 3 |
| 55 | Decorated | **Decor** | **Decor** | 4 | 4 |
| 56 | Decorating | **Decor** | **Decor** | 5 | 5 |
| 57 | Employed | **Employ** | **Employ** | 2 | 2 |
| 58 | Employing | **Employ** | **Employ** | 3 | 3 |
| 59 | Explained | **Explain** | **Explain** | 2 | 2 |
| 60 | Explaining | **Explain** | **Explain** | 3 | 3 |
| 61 | Finished | **Finish** | **Finish** | 2 | 2 |
| 62 | Finishing | **Finish** | **Finish** | 3 | 3 |
| 63 | Forbidden | Forbidden | Forbid | 0 | 3 |
| 64 | Forbidding | **Forbid** | **Forbid** | 4 | 4 |
| 65 | Gathered | **Gather** | **Gather** | 2 | 2 |
| 66 | Gathering | **Gather** | **Gather** | 3 | 3 |
| 67 | Improved | **Improv** | **Improv** | 2 | 2 |
| 68 | Improving | **Improv** | **Improv** | 3 | 3 |
| 69 | Laughed | **Laugh** | **Laugh** | 2 | 2 |
| 70 | Laughing | **Laugh** | **Laugh** | 3 | 3 |
| 71 | Listened | **Listen** | **Listen** | 2 | 2 |
| 72 | Listening | **Listen** | **Listen** | 3 | 3 |
| 73 | Mended | Mend | Mende | 2 | 1 |
| 74 | Mending | Mend | Mending | 3 | 0 |
| 75 | Nipped | Nip | Nippe | 3 | 1 |
| 76 | Nipping | Nip | Nipp | 4 | 3 |
| 77 | Plucked | **Pluck** | **Pluck** | 2 | 2 |
| 78 | Plucking | **Pluck** | **Pluck** | 3 | 3 |
| 79 | Preached | **Preach** | **Preach** | 2 | 2 |
| 80 | Preaching | **Preach** | **Preach** | 3 | 3 |
| 81 | Enormously | **Enorm** | **Enorm** | 5 | 5 |
| 82 | Monthly | **Month** | **Month** | 2 | 2 |
| 83 | Solemnly | **Solemn** | **Solemn** | 2 | 2 |
| 84 | Abnormally | Abnorm | Abnormal | 4 | 2 |
| 85 | Diligently | Dilig | Diligen | 5 | 3 |
| 86 | Jubilantly | **Jubil** | **Jubil** | 5 | 5 |
| 87 | Frightfully | **Fright** | **Fright** | 5 | 5 |
| 88 | Swiftly | **Swift** | **Swift** | 2 | 2 |
| 89 | Miserable | **Miser** | **Miser** | 4 | 4 |
| 90 | Thankfully | Thank | Thankful | 5 | 2 |
| 91 | Blissfully | Bliss | Blissful | 5 | 2 |
| 92 | Reluctantly | Reluct | Reluctan | 5 | 3 |
| 93 | Viciously | **Vicious** | **Vicious** | 2 | 2 |
| 94 | Wonderfully | **Wonder** | **Wonder** | 5 | 5 |
| 95 | Hopelessly | **Hopeless** | **Hopeless** | 2 | 2 |
| 96 | Briskly | Brisk | Briskly | 2 | 0 |
| 97 | Delightfully | **Delight** | **Delight** | 5 | 5 |
| 98 | Anxiously | **Anxious** | **Anxious** | 2 | 2 |
| 99 | Obnoxiously | Obnoxi | Obnoxious | 5 | 2 |
| 100 | Inwardly | **Inward** | **Inward** | 2 | 2 |

We hypothesize that if our stemmer is as efficient as that of Porter's, then the distributions of Levenshtein distances between a word and its stripped stem tend to be same. For a given word, we have two treatments in hand, Porter's algorithm and our N-gram procedure. Thus for a given word, we have a pair of LDs. For comparison, we set the null hypothesis $H_0$ that no differences exist between the two stemmers.

Because the two sets of measures can be considered as two measures associated to the same sample, we decide to employ a statistical test for paired samples. In particular, a nonparametric statistical test, the *Wilcoxon signed-rank test* is employed by us, since there was no evidence about the distribution of these distances. The Wilcoxon test is based on two paired series ($x_i$, $y_i$) of $N$ observed values, one series for each out of the observed variables $X$, $Y$ to be compared. The absolute differences $d_i=abs\ (x_i-y_i)$ and $sign\ (d_i)$ of differences are to be computed next. Then the absolute values are ranked, discarding zeros. If $N_r$ is the reduced sample size, the final test statistic is $W = |\sum_i^{N_r}[sign(d_i) * rank(|d_i|)]|$, which is approximated by a Normal variable for large N.

## 4. Experimental Results:

On conducting the Wilcoxon test over our sample data, we get the $p$ value of 0.54. As $p>0.05$, fail to reject the null hypothesis $H_0$. Thus, we have strong evidence that our N-gram Stemmer is not inferior to Porter's Stemmer.

## 5. Conclusions:

We endeavoured to modify a statistical stemming technique and came up with one whose results are comparable with a well known linguistic stemmer. Our N-gram stemmer is capable to deal any language for which N-gram corpus frequencies are available. As a language neutral approach is always preferable over a technique in which a linguistic knowledge or analysis is prerequisite, our results seem to be very exciting and promising.